\documentclass[prl,aps,nofootinbib,11pt]{revtex4}
\usepackage{graphicx,epsfig}
\usepackage{graphicx,color}

\begin{document}
\title{Decipher the short-distance component of $X(3872)$ in $B_c$ decays}

\author{Wei Wang$^{1,2}$\footnote{Email:wei.wang@sjtu.edu.cn} and Qiang Zhao$^3$\footnote{Email:zhaoq@ihep.ac.cn} }

\affiliation{
$^1$ INPAC, Shanghai Key Laboratory for Particle Physics and Cosmology, Department of Physics and Astronomy, Shanghai Jiao-Tong University, Shanghai, 200240,   China\\
$^2$
State Key Laboratory of Theoretical Physics, Institute of Theoretical Physics, Chinese Academy of Sciences, Beijing 100190, China\\
$^3$ Institute of High Energy Physics and Theoretical Physics Center for Science Facilities,\\
        Chinese Academy of Sciences, Beijing 100049, China }

\begin{abstract}
A foremost  task in  understanding the nature of the $X(3872)$  involves the
discrimination of the two-quark  and multiquark configurations.  In this work, we propose a method to probe the short-distance component  of the $X(3872)$ by measuring the ratio between the $B_c$ semileptonic and nonleptonic decays into the $X(3872)$.
We demonstrate that if the  $X(3872)$ production mechanism  is through  the $\bar cc$ component,  the ratios would  be universal  and  could  be reliably  predicted in theory.  Measurements of these ratios at   LHC and  the next-generation  electron-positron colliders are capable of validating/invalidating this  production mechanism and providing deeper insights into the nature of the $X(3872)$.
\end{abstract}
\maketitle


Thanks to the unprecedented data samples accumulated by the  two $B$ factories and high energy hadron colliders,  dramatic progresses have been made in the study of  hadron
spectroscopy in the past decades. In particular, in  the  heavy
quarkonium sector, a number of  unexpected  resonance-like structures have been discovered  at these
experimental facilities,  among which the $X(3872)$ is one most notable example~\cite{Swanson:2006st,Godfrey:2008nc,Eichten:2007qx,Brambilla:2010cs}.

The $X(3872)$ was first discovered  in the exclusive decay
$B^{\pm}\rightarrow K^{\pm}X(3872)\rightarrow K^{\pm}\pi^+\pi^- J/\psi$ by   Belle Collaboration  at
the $e^+e^-$ collider located at KEK~\cite{Choi:2003ue} and then confirmed
by the BaBar Collaboration in the same channel~\cite{Aubert:2004ns}.
This meson has also been observed in the  high energy hadron-hadron colliders  at the Tevatron~\cite{Acosta:2003zx,Abazov:2004kp,Abulencia:2005zc,Aaltonen:2009vj} and
LHC~\cite{Chatrchyan:2013cld,Aaij:2013zoa,Aaij:2015eva}.   Based on the  data corresponding
to an integrated luminosity of $3.0 fb^{-1}$ of proton-proton collisions, the LHCb collaboration has performed an angular analysis of the $X(3872)$ decay   and found the quantum numbers $J^{PC}=1^{++}$~\cite{Aaij:2015eva}.  The $X(3872)$ meson is peculiar in several
aspects, and  its nature is still not well-understood.  Its  width  is  tiny
compared to typical hadronic widths and only an upper bound has been set to date:
$\Gamma<1.2$~MeV~\cite{Agashe:2014kda}. The mass lies in the vicinity to the  $D^0\bar D^{*0}$ threshold,
$M_{X(3872)}-M_{D^0}-M_{D^{*0} }=(-0.12\pm0.24)$~MeV~\cite{Lees:2013dja},
which leads  to speculations of the $X(3872)$  as a hadronic
molecule: a $D^0\overline D^{*0}$ loosely bound state~\cite{Tornqvist:2004qy} or   a virtual state~\cite{Hanhart:2007yq}.  Meanwhile other non-charmonium explanations  were also proposed in the literature,
such as $\bar ccg$ hybrid meson~\cite{Li:2004sta},
glueball~\cite{Seth:2004zb}, and a compact tetraquark state as the diquark cluster~\cite{Maiani:2004vq}.

A very important task in  understanding the nature of the $X(3872)$  involves the
discrimination of a two-quark configuration as the $\chi_{c1}(2P)$,  a compact multiquark configuration and a
hadronic molecule~\cite{Close:2003sg,Meng:2007cx,Li:2009ad,Butenschoen:2013pxa,Meng:2013gga,Chen:2013upa}.  But in fact, there are few experimental processes which can provide a clean discrimination among these descriptions, which  makes the situation  obscure. In this work, we propose an approach  that is able to directly examine the structure of the $X(3872)$ at short distance, and probe its short-distance component in the $B_c$ semileptonic and nonlepotnic decays.  As we will show later, the ratios between   branching fractions in the $B_c$ semileptonic and nonlepotnic decays are almost universal for different polarizations of the $X(3872)$ and can be reliably predicted in theory if the production is dominated by the $\bar cc$. The decay modes include the semileptonic  $B_c^-\to X(3872)\ell^-\bar\nu$, and the $B_c^-\to X(3872) \rho^-,X(3872)a_1^-(1260)$ decays. The light meson in the final state can also be replaced by the strange mesons $K^*(892)$, $K_1(1270)$ or $K_1(1400)$, with the cost of the reduced branching fractions due to the suppressed CKM matrix element $|V_{us}|$.

\begin{figure}\begin{center}
\includegraphics[scale=0.5]{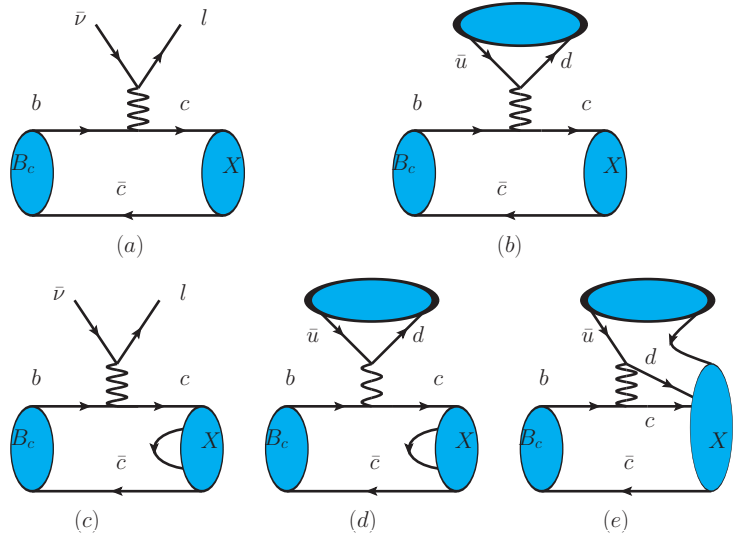}
\caption{Feynman diagrams for the $B_c$ decays into the $X(3872)$: the upper two diagrams for the $\bar cc$ structure, and the lower ones for the four-quark assignments.  } \label{fig:feyman}
\end{center}
\end{figure}

Hereafter we will use the abbreviation $X$ to denote the $X(3872)$ for the sake of simplicity.
Feyman diagrams for the semileptonic and nonleptonic $B_c\to X(3872)$ transitions are  given  in Fig.~\ref{fig:feyman}: the upper two diagrams correspond to the $\bar cc$ configuration, while the lower ones correspond to the four-quark case.

We will first discuss the implications from  the $\bar cc$ component and start with the semileptonic decay mode, in which the $B_c\to X$ transition  induced by the vector and
axial-vector currents is parametrized by:
 \begin{eqnarray}
  &&\langle X|\bar c\gamma_{\mu}\gamma_5 b|B_c^-\rangle
    = -\frac{2iA(q^2)}{m_{B_c}-m_{X}}\epsilon_{\mu\nu\rho\sigma}
     \epsilon^{*{\nu}}p_{B_c}^{\rho}p_{{X}}^{\sigma},
     \nonumber\\
  &&\langle X|\bar c\gamma_{\mu}b| B_c^-\rangle
    = -2m_{X} V_0(q^2)\frac{\epsilon^*\cdot q}{q^2}q_{\mu}
    -   (m_{B_c}-m_{X})V_1(q^2)\left[\epsilon^*_{\mu}
    -\frac{\epsilon^*\cdot q}{q^2}q_{\mu} \right] \nonumber\\
    &&\;\;\;\;+V_2(q^2)\frac{\epsilon^*\cdot q}{m_{B_c}-m_{X}}
     \left[ P_{\mu}-\frac{m_{B_c}^2-m_{X}^2}{q^2}q_{\mu} \right],
 \end{eqnarray}
 with $P=p_{B_c}+p_{X}$,  $q=p_{B_c}-p_{X}$, and $\epsilon^{0123}=+1$.


We will consider the $\ell=e,\mu$ and thus neglect the lepton mass.  The differential decay width for  the $B_c^-\to X\ell^-\bar\nu_{\ell}$ is  given  as
\begin{eqnarray}
&& \frac{d\Gamma(B_c^-\to X\ell^-\bar\nu_{\ell})}{dq^2  }
 =  \frac{\sqrt{\lambda(q^2)} q^2 G_F^2|V_{cb}|^2}{ 192 \pi^3m_{B_c}^3 }   \times [ |{\cal A}^1_{0}(q^2)|^2  + |{\cal A}^1_{\perp}(q^2)|^2+|{\cal A}^1_{||}(q^2)|^2], \label{eq:dgammadq2semileptonic}
\end{eqnarray}
where $|V_{cb}|$ is the CKM matrix element and $G_F$ is the Fermi constant. The  $q^2$ as the lepton pair invariant mass square and
\begin{eqnarray}
\lambda(q^2)=(m_{B_c}^2+m_{X}^2-q^2)^2-4m_{B_c}^2m_{X}^2. \nonumber
\end{eqnarray}
The  polarised decay amplitudes are  defined  as
\begin{eqnarray}
 {\cal A}^1_0(q^2)&=&  \frac{1}{2m_{X}\sqrt {q^2}}\bigg[-\frac{\lambda(q^2)}{m_{B_c}-m_{X}}V_2(q^2)
\;\;+(m_{B_c}^2-m_{X}^2-q^2)(m_{B_c}-m_{X})V_1(q^2)
 \bigg], \nonumber\\
 {\cal A}^1_\pm(q^2)
 &=&    (m_{B_c}-m_{X})V_1(q^2)\mp \frac{\sqrt {\lambda(q^2)}}{m_{B_c}-m_{X}}A(q^2),\nonumber\\
 {\cal A}^1_{\perp/||}(q^2)&=&\frac{1}{\sqrt 2}[ {\cal A}^1_{+}(q^2) \mp  {\cal A}^1_{-}(q^2)].
\end{eqnarray}


After the integration of the off-shell $W$-boson, the effective Hamiltonian for the $b\to c\bar u d$ transition  is given as
\begin{eqnarray}
 {\cal H}_{\rm eff} &=& \frac{G_{F}}{\sqrt{2}}V_{cb} V_{ud}^{*}
     \bigg\{
     C_{1} O_1
  +  C_{2} O_2 \bigg\},
 \label{eq:hamiltonian01}
 \end{eqnarray}
 where $C_1$ and $C_2$ are   Wilson coefficients for the operators $O_{1}$ and $O_2$.  $V_{cb}, V_{ud}$ are the CKM matrix elements.
If the $X(3872)$ is composed of the $\bar cc$, the above effective Hamiltonian leads to
\begin{eqnarray}
 &&\Gamma(B_c^-\to X\rho^-)= \frac{|\vec p|}{8\pi m_{B_c}^2}   \left|\frac{G_F}{\sqrt 2}V_{cb}V_{ud}^* a_1  f_{\rho}m_{\rho}\right|^2
 \times [ |{\cal A}^1_{0}(m_{\rho}^2)|^2  + |{\cal A}^1_{\perp}(m_{\rho}^2)|^2+|{\cal A}^1_{||}(m_{\rho}^2)|^2],\label{eq:dgammadq2nonleptonic}
\end{eqnarray}
where $a_1=C_1+C_2/3$ and $|\vec p|$ is the three momentum of the $X(3872)$ in the $B_c$ rest frame.
The $f_{\rho}$ and $m_{\rho}$ is the $\rho$ meson decay constant and mass, respectively.
In deriving the above formulas, we have assumed the factorization theorem, which can be proved at leading power in $1/m_{b}$ using soft-collinear-effective theory~\cite{Bauer:2000ew,Bauer:2000yr} similar with the proof for the $\overline B^0\to D^+\pi^-$ channel~\cite{Bauer:2001cu}.


From Eq.~(\ref{eq:dgammadq2semileptonic}) and Eq.~(\ref{eq:dgammadq2nonleptonic}), we can see most  hadronic effects will  cancel if we consider the ratios of branching fractions:
\begin{eqnarray}
  R_{i}(\rho)=\int_{(m_{\rho}-\delta)^2}^{(m_\rho+\delta)^2}dq^2\frac{d{\cal B}(B_c^-\to X_{i} \ell^-\bar\nu_{\ell})}{dq^2}  \frac{1}{{\cal B}(B_c^-\to X_{i}\rho^-) }. \label{eq:ratio_rho}
\end{eqnarray}
In Eq.~(\ref{eq:ratio_rho}) the subscript $i$ denotes the polarization, with $i=0,\perp,||$ or $i=0,\pm1$, or the total decay width. In order to access the semileptonic decay modes on the experimental side, we have limited the analysis to the mass region around the $\rho$ meson mass, with the parameter $\delta$ charactering the size of the bin.


In the case that the production of the $X(3872)$ is dominated by the $\bar cc$, the above ratios are predicted as
\begin{eqnarray}
R_0(\rho) =  (10.9\pm0.1)\times 10^{-3},\nonumber\\
 R_{\perp}(\rho)=   (11.1\pm0.1)\times 10^{-3},\nonumber\\
 R_{||}(\rho)=  (11.1\pm0.1)\times 10^{-3},\nonumber\\
R_{\rm total}(\rho)=  (10.9\pm0.1)\times 10^{-3},  \label{eq:result_rho}
\end{eqnarray}
where for illustration we have used $\delta=0.1$ GeV which is at the same order of magnitude with the $\rho$ meson width.   Choosing a different $\delta$ will be   similar.  The errors  given in Eq.~(\ref{eq:result_rho}) arise from transition form factors. For an estimate, we have used the constant form factors, the calculation  in covariant light-front approach~\cite{Wang:2007sxa}, and  light-cone sum rules~\cite{Wang:2007fs}.
In the numerics, we have also used $m_{\rho}=0.77526$ GeV and $a_1=1.07$~\cite{hep-ph/9512380}.  The $f_{\rho}=(209.4\pm0.4)$ MeV is extracted from the data on $\tau\to \rho^-\nu_\tau$ decay~\cite{Agashe:2014kda}.
As we can see the above ratios are universal and stable against the hadronic uncertainties.

The $\rho^-$ meson mainly decays into  the $\pi^-\pi^0$, in which the neutral pion may be difficult to reconstruct. In this case, it may be more advantageous to consider the $a_1(1260)$ which decays into three pions. In fact, the decay of $B_c\to J/\psi \pi^+\pi^-\pi^-$ has been  observed by LHCb~\cite{LHCb:2012ag} and CMS~\cite{Khachatryan:2014nfa} collaboration, in which  the $a_1(1260)$ provides the dominant contribution.  For our purpose, we can  similarly  define
\begin{eqnarray}
  R_{i}(a_1)=\int_{(m_{a_1}-\delta)^2}^{(m_{a_1}+\delta)^2}dq^2\frac{d{\cal B}(B_c^-\to X_{i} \ell^-\bar\nu)}{dq^2}  \frac{1}{{\cal B}(B_c^-\to X_{i}a_1^-) }.  \label{eq:ratio_a1}
\end{eqnarray}
Again if the production  is mostly through the $\bar cc$, the above ratios are predicted as
\begin{eqnarray}
R_0(a_1) =  (13.5\pm0.1\pm1.1)\times 10^{-3},\nonumber\\
 R_{\perp}(a_1)=   (13.5\pm0.1\pm1.1)\times 10^{-3},\nonumber\\
 R_{||}(a_1)= (13.5\pm0.1\pm1.1)\times 10^{-3},\nonumber\\
R_{\rm total}(a_1)=  (13.5\pm0.1\pm1.1)\times 10^{-3}. \label{eq:result_a1}
\end{eqnarray}
The first errors originate   from the $B_c\to X$ form factors and the second ones are from the $f_{a_1}$ for which we have used $ f_{a_1}= (238\pm 10){\rm MeV}$~\cite{Yang:2007zt}.  This sizable error   is reducible using  the experimental  data on $\tau\to a_1^-(1260)\nu_\tau$.

{ One can also use $K^*(892)$ or $K_1(1270)/K_1(1400)$ to tag the production mechanism for the $X(3872)$.  The price to pay is that the decay amplitude is proportional to  the smaller CKM matrix element $V_{us}$ compared to the $V_{ud}$ in the associated production of $\rho$ and $a_1(1260)$. }  For the $K^*(892)$ final state, we have
 \begin{eqnarray}
R_0(K^*) =  (0.245\pm0.001\pm0.014),\nonumber\\
 R_{\perp}(K^*)=   (0.247\pm0.001\pm0.014),\nonumber\\
 R_{||}(K^*)= (0.249\pm0.001\pm0.014),\nonumber\\
R_{\rm total}(K^*)= (0.246\pm0.001\pm0.014) \label{eq:result_Kstar}
\end{eqnarray}
where again the errors are from the $B_c\to X$ transition form factors and the $K^*$ decay constant extracted from the $\tau\to K^*\nu$: $f_{K^*}= (205\pm6)$MeV.

Based on the huge amount of  data samples, the LHC experiment  is  playing  an important role in the study of hadron exotics.  The LHCb collaboration has measured the $B$ decays into the $X(3872)$ and determined its quantum numbers~\cite{Aaij:2013zoa,Aaij:2015eva}.
Based on the  $1.0fb^{-1}$ data at the center-of-mass (c.m.) energy of 7 TeV, the LHCb collaboration is also able to extract the ratio of $B_c^+$ branching fractions to $J/\psi\pi^+$ and $J/\psi\mu^+\nu_\mu$~\cite{Aaij:2014jxa}.
For the nonleptonic $B_c$ decays into the $X(3872)$,  a theoretical  estimate of their branching fractions is given in Ref.~\cite{Wang:2007sxa}
\begin{eqnarray}
{\cal B}(B_c^-\to X\rho^-)= (5.0^{+2.0}_{-1.7})\times 10^{-3},\nonumber\\
{\cal B}(B_c^-\to XK^{*-})= (2.9^{+1.1}_{-1.0})\times 10^{-4},
\end{eqnarray}
where the $X(3872)$ is treated as a $\chi_{c1}(2P)$ state. In the future the data sample will be increased by at least one order of magnitude, and thus it is very  likely for the LHCb to observe the $B_c$ decays into the $X(3872)$ due to the sizable branching fractions. In addition,
the experimental prospect at the next-generation electron-positron collider is also promising, for instance, the CEPC will produce about $10^{11}$ $b\bar b$ events  at the c.m. energy $\sqrt s= m_Z$~\cite{CEPC_preCDR}.

The independence on hadronic effects of the above ratios is  evident for the processes of Fig.~\ref{fig:feyman} (a) and (b). If the $X(3872)$ is composed of four-quarks at  short-distance, as either a compact tetraquark  or hadronic molecule, the situation will be different. In this case,  the production Feynman diagrams  are demonstrated in Fig.~\ref{fig:feyman} (c),(d), and (e). Since a pair of light-quarks are produced at short-distance compared to Fig.~\ref{fig:feyman} (a) and (b), the production rates will be greatly suppressed by the strong coupling constant and powers of the $1/m_b$, which can lead to  very small branching fractions for the $B_c\to X$ transition. Moreover, there will be sources  spoiling  the relation for the ratios as given in Fig.~\ref{fig:feyman} (e).  So   a sizable production rate and the agreement between the data and predictions on ratios between  branching fractions of $B_c$ decays will imply the presence of a $\bar{c}c$ core within the $X(3872)$.   Alternatively, a mismatch of the predicted ratios will clearly  indicate the short-distance non-$\bar cc$ component in the $X(3872)$.

To summarize, although the $X(3872)$ meson has been well established in experiment, its nature is still under debate due to prescriptions from different scenarios.  In this work, we propose a method to explore its short-distance $\bar cc$ component in the semileptonic and nonleptonic $B_c$ decays by measuring the production ratios of branching fractions. We  demonstrate that these ratios are almost universal and can be reliably predicted if there exist a $\bar cc$ component within the $X(3872)$.  These predictions could  be directly tested by  the measurements in the future.  Significant deviations from the   results in Eqs.~(\ref{eq:result_rho}), (\ref{eq:result_a1})  and (\ref{eq:result_Kstar}) would be a clear signal for the non-standard charmonium structure  at the short distance. With the large amount of data in the future, we would expect that the above predictions can be examined and much deeper insights into the nature of the $X(3872)$ can be achieved.

 {\it Acknowledgements:}
The authors are very  grateful to Ahmed Ali, Feng-Kun Guo,   Xiao-Gang He and Qian Wang  for  enlightening discussions. This work is supported in part  by National  Natural  Science Foundation of China under Grant  No.11575110 and 11425525,  Natural  Science Foundation of Shanghai under Grant  No.11DZ2260700, 15DZ2272100 and No.15ZR1423100,   the Sino-German CRC 110 ``Symmetries and
the Emergence of Structure in QCD" (NSFC Grant No. 11261130311), the Open Project Program of State Key Laboratory of Theoretical Physics, Institute of Theoretical Physics, Chinese  Academy of Sciences, China (No.Y5KF111CJ1), and  Scientific Research Foundation for  Returned Overseas Chinese Scholars, State Education Ministry.


\end{document}